# The influence of system dynamics on the frictional resistance: insights from a discrete model

Robbin Wetter, Valentin L. Popov

*Technische Universität Berlin, 10623 Berlin, Germany*
r.wetter@tu-berlin.de, 0049-30-314 22154

*Abstract:* In order to examine the influence of system dynamics on sliding friction, we introduce the so-called micro-walking machine. This model consists of a rigid body with a number of elastic contact spots that is pulled by a constantly moving base. The system slides with dry friction on a rigid substrate. The kinematic coupling of the rotation and the translation of the rigid body results in varying normal and tangential forces at the contact spots. For certain parameter ranges this leads to self-excited oscillations in the vertical direction. A particular dynamic mode occurs which is characterized by a strong correlation between low or even zero normal forces and a fast forward motion. This effect is referred to as micro-walking. In addition to an experimental rig we use numerical integration and an extensive parameter study for the analysis. In theory, the reduction of the frictional resistance reaches up to 98%. These results are confirmed by the experiments where the maximal reduction was 73%.

Our model shows that micro-vibrations play an important role for the dynamic influences on the frictional resistance of systems that exhibit apparently smooth sliding. The identification of the critical parameter range enables the systematic control of frictional resistance through the adjustment of attributes such as geometry and stiffness. In addition, it is possible to deduce guidelines for how tribological test rigs should be designed in order to get reliable results.

## 1. Introduction

Frictional resistance is a crucial parameter in technical, seismological and even biological systems [1]. The interaction of two solids is often modelled with the well-known Coulomb's law of dry friction. This model goes back to the works of Amontons [2] and Coulomb [3] and gives a proportional relation between the normal load $F_N$ and the tangential load $F_T$ of the contact. As long as $F_T$ falls below the maximum holding force:

$$F_{T,\max} = \mu_s F_N \qquad (1)$$

the system remains at rest. Otherwise, a relative movement of the two bodies occurs and both are subjected to the resistance force:

$$F_R = \mu_k F_N . \qquad (2)$$

The two constants of proportionality denote the so called static $\mu_s$ and dynamic $\mu_k$ coefficient of friction. One explanation for the proportionality of Coulomb's law lies in the contact properties of rough surfaces. According to the theory by Bowden and Tabor [4] rough surfaces consists of numerous small hills named asperities. Thus, the two bodies only touch in a few contact spots where, in case of dry metals, cold-weld junctions are formed. Assuming that the number of contact spots is proportional to the normal load, the coefficient of friction equals the ratio of the traction necessary to shear the cold-weld junctions and the penetration hardness. This explains also why experimentally, the static coefficient $\mu_s$ often exceeds the kinetic one $\mu_k$ since the metallic junctions become stronger after the surfaces have been in stationary contact for some time [5].

The coefficients are determined experimentally and are classified for a large number of combinations of contacting materials. This often leads to the widespread misconception of Coulomb's law of dry friction according to which the coefficients of friction are true constants that only depend on intrinsic properties of the materials in contact [6]. In fact they depend on small scale effects in the contact and are influenced by a various time and environmental conditions for both lubricated and dry contacts. It should be emphasized that Coulomb was already fully aware of these many influences and also presented a lot of experimental work on this subject [7]. In particular, the material combination, the contact geometry, chemical reactions, the temperature and the normal force itself all play an important role [6]. The static coefficient is also affected by the history of the contact, as evidenced by the fact that it increases with time [5, 8].

Furthermore, the kinetic coefficient is strongly affected by dynamic influences as sliding velocity and vibrations [9-13]. This is a particular problem as it affects every measurement apparatus, i.e. tribometer, which is used for determination of the coefficients. On the one hand, this limits the value of tabulated coefficients as the influence of the tribometer used in the experiment remains vague [6, 14]. On the other hand, this is a possible explanation for difficulties on the reproducibility of coefficients that are determined with different tribometers under otherwise similar conditions [15]. Dynamic influences on the frictional resistance include for instance damping, self-excited and externally excited vibrations, the velocity, the contact stiffness and the interaction between the structural dynamic behaviour and the excitation by the frictional force [6]. In particular the impact of system dynamics on the interface response of frictional contacts is a crucial factor. This is mainly caused by a coupling between the in-plane and out of plane vibrations in sliding systems [16].

An important work is the experiment of Tolstoi, who studied the interaction of the tangential and normal degree of freedom [17]. He found that tangential slip events are accompanied by an upward movement of the contacting body and that damping increases the frictional resistance. In addition, he used piezo actuators to induce externally excited vibrations in the normal direction. Through resonance effects, the frictional resistance was reduced between 35% and 85%. Godfrey used electrical resistance measurements and showed that this effect is due to a reduction of the metal-to-metal contact zone induced by decreasing loads. The reduction in the coefficient of friction was therefore characterized as apparent [18]. Polycarpou and Soom used linear dynamic models to compute the normal motion of a lubricated sliding system. They concluded that the instantaneous normal separation significantly affects the friction force. A good representation of the dynamics of the sliding system in the normal direction is therefore very important [19-21]. Several authors added a rotational degree of freedom. By this, they considered that the kinematic coupling between normal, tangential and rotational motion leads to varying forces and moments what in turn influences the friction [15, 22, 23]. Twozydlo et al. modelled a typical pin-on-disk apparatus consisting of rigid bodies with elastic connections. They showed that coupling between rotational and normal modes induces self-excited oscillations. In combination with high-frequency stick-slip motion, these oscillations reduce the apparent kinetic coefficient of friction. A particular pin-on-disk experimental set-up gave good qualitative and quantitative correlation with numerical results [24]. Adams examined a beam that is in frictional contact with an elastic foundation [25]. In this model the interplay of the friction force and moments at the interface and the bending of the beam lead to instabilities that increase with increasing coefficient of friction. This self-excited oscillation leads to a partial loss of contact and a stick-slip motion.

Despite discrete lumped parameter models, many authors considered interface waves with separation between contacting half-spaces. Caminou and Dundurs investigated the so-called carpet-fold motion, which gives the possibility of a sliding motion between two identical bodies without interface slipping [26], i.e. without frictional resistance. Adams examined elastic half-spaces in dry contact and concluded that the deformation along the interface and slip wave propagation is a potential destabilizing mechanism for steady sliding [27]. Although his model uses a constant coefficient of friction without distinction between the static and kinetic case, he demonstrated that interface slip waves may be responsible for the apparent velocity-dependence of friction coefficient measurements [28]. The apparent coefficient of friction can decrease with increasing sliding speed due to the carpet-fold motion.

Another scenario is an oscillating force that causes micro-slip to occur in different contact areas for each loading cycle. Mugadu et al. introduced a rigid flat punch loaded by a constant tangential force insufficient to cause gross sliding [29]. A constant normal force that moves backwards and forwards on the punch causes micro-slip that alternately migrates in from one side of the contact. This induces a rocking motion of the punch such that it walks across an elastic half plane by a constant increment in each loading cycle, even though the global friction limit is not exceeded at any time. Nowell used numerical methods to also study the case for a normal force being strong enough to alternately completely lift the punch from the one end [30].

Taken together, it shows that vibrations in the normal direction decrease the frictional resistance of mechanical systems. This effect has long been known and is also applied in technical systems [17, 31]. Furthermore friction and dynamic coupling effects can lead to instabilities [27, 28]. This raises the question to what extent these self-excited vibrations can lead to a self-excited reduction of the macroscopic frictional resistance. In case that the corresponding amplitudes are of a microscopic character, they could be superimposed to an apparently smooth sliding motion of the system while being undetected.

In the present paper we introduce a model that exhibits this kind of behaviour. We use both experiments and simulations to examine the underlying dynamic effects and to show which parameters contribute mostly to this. It turns out that the frictional resistance of our particular system almost vanishes for certain parameter ranges. Thus, in the limits of the proposed model, we can deduce recommendations for the minimization of dynamic influences in tribological test rigs or, vice versa, for the reduction of frictional resistance in technical systems.

The paper is structured as follows. In section 2 we propose a thought experiment for a possible reduction effect and introduce the corresponding model. In section 3 we examine the parameters of influence and give an explanation for the friction minimizing effects. Section 4 describes the experimental setting and gives a comparison of theoretical and experimental results. Finally a conclusion is provided in section 5.

## 2. Modelling and methods

A potential basic system that captures important dynamic effects that enable a significant reduction of the apparent frictional resistance could be as follows. Assume an elastic specimen which is pressed on a rigid substrate by the macroscopic normal force $F_N$ and is additionally loaded in the tangential direction with the macroscopic tangential force $F_T$, such that it slides with a constant velocity $v_0$ as shown in Fig. 1 (a). As in real contacts, specimen and substrate will not touch over the entire apparent contact surface but in several small contact zones [4]. Each contact zone will exhibit its own stress distribution in the normal and tangential direction that is represented by a normal and a tangential force $F_{Ni}$ and $F_{Ti}$ as depicted in Fig. 1 (b).

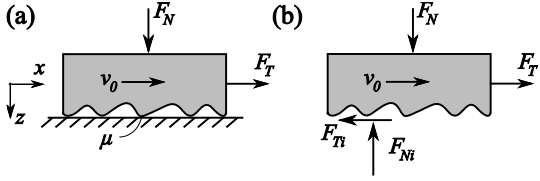

**Fig. 1** Elastic specimen that slides with constant velocity on a rigid substrate (**a**). Free body diagram with normal and tangential forces of one contact spot (**b**)

We restrict ourselves to the case of dry friction and assume that Coulomb's law with a constant coefficient $\mu_s = \mu_k = \mu$ applies in the contact spots. Thus, we neglect any distinction between static and kinetic friction and the variation of the latter with sliding speed as in [15, 28]. The coefficient $\mu$ reflects all microscopic influences of the surface that depend on the material pairing. According to the microscopic sticking condition a specific spot sticks whenever:

$$F_{Ti} \leq \mu F_{Ni}. \quad (3)$$

Otherwise the contact spot slips and is subjected to a microscopic resistance force:

$$F_{Ri} = \mu F_{Ni}. \quad (4)$$

Due to the kinematic coupling between the normal and tangential translation and the rotation, the forces in the contact spots will vary in time. This effect significantly influences the apparent frictional resistance [15, 23] as can be explained as follows. Assuming that there are overall $n$ similarly loaded contact spots we can define the theoretical average normal force acting in one contact spot as:

$$\overline{F}_{Ni} = \frac{F_N}{n}. \quad (5)$$

Due to the elasticity of the specimen, the actual status of a specific contact spot, i.e. sticking or slipping, is independent from the actual status of the rigid body. Thus, the rigid body can move in the tangential direction while a single spot is sticking. In addition, we assume that the distance between the spots is sufficiently large so that the spots are independent. The relative displacement between substrate and bulk body would thus only change the elastic deformation in the vicinity of a sticking spot. Under these assumptions the system will be able to micro-walk on the substrate. In this case micro-walking describes a particular behaviour of the contact spots:

- the spots only slip whenever their actual normal force $F_{Ni}$ falls below $\overline{F}_{Ni}$
- the spots stick whenever their actual normal force $F_{Ni}$ exceeds $\overline{F}_{Ni}$

As a result, the average resistance force of contact spots $\overline{F}_{Ri} = \langle F_{Ri} \rangle$, where $\langle . \rangle$ denotes the time average, falls below its theoretical value calculated with (4) and (5):

$$\overline{F}_{Ri} < \mu \overline{F}_{Ni}. \quad (6)$$

In consequence, the overall resistance force of the system, which is the sum over all of the $n$ contact spots, falls below the overall theoretical value as well:

$$F_R = \sum_{i=1}^{n} \overline{F}_{Ri} < \sum_{i=1}^{n} \mu \overline{F}_{Ni} = \mu F_N. \quad (7)$$

This leads to the definition of the effective respectively apparent coefficient of friction $\mu_e$:

$$\mu_e = \frac{F_R}{\mu F_N}. \quad (8)$$

There is a strong analogy to the ratcheting case described in [29] where one side of a single contact slips while the other sticks. This micro-slip accumulates to a rigid body motion. Hence, ratcheting decreases the tangential load that is needed for gross sliding i.e. decreases the apparent static and kinetic friction.

An experimental analogy of the dynamic mechanism of friction reduction is given by the work of Tolstoi [17]. In this case, increasing sliding speed increases the normal oscillations of the sliding body. Due to the non-linearity between indentation and normal force the vibrations of the slider are highly asymmetric. Consequently, an increase of the normal oscillation amplitude decreases the mean value of indentation and the contact radius of the contact spots during sliding. As a result, the friction force decreases as well [15]. Our model extends this concept, as we also consider the spatial variation of normal and tangential forces and stick and slip zones.

### 2.1. Discrete model

The simplest implementation of a system that captures the aforementioned effects is a plane specimen that has only two contact spots at the edges, as depicted in Fig. 2.

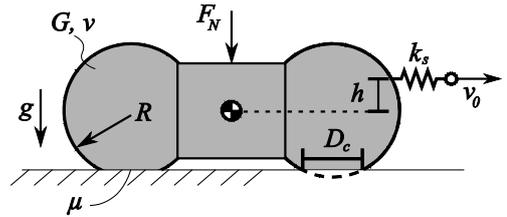

**Fig. 2** Specimen consisting of two spherical shaped edges pulled by a constantly moving base

The specimen is unloaded in the normal direction except for the force of gravity of the bulk body $mg$ which refers to the macroscopic normal force $F_N$. Here $m$ denotes the mass of the specimen and $g$ is the gravitational acceleration. The tangential force $F_T$ is applied by a constantly moving base that is connected via a spring with stiffness $k_s$. As the base moves with constant velocity $v_0$ the mean velocity of the specimen $\overline{v}$ corresponds to $v_0$. This type of excitation models a nominally steady motion of the system. Thus, any vibrations that may occur are not directly caused by the excitation but rather by self-excited oscillations. We use the separation of space scales principle for a further simplification. According to this the space scales contributing to the elastic and kinetic energy of a mechanical system with friction can, under most conditions, be separated in the following way [32]:

- The elastic energy is a *local* quantity which only depends on the conditions in the contact region
- The kinetic energy is a *non-local* quantity which can be assumed to equal to the kinetic energy of the rigid bulk as a whole

This enables a further simplification, such that there remains a rigid body with mass $m$, moment of inertia $\Theta$, height $2a$ and length $2b$ that consists of two elastic contact spots as shown in Fig. 3.

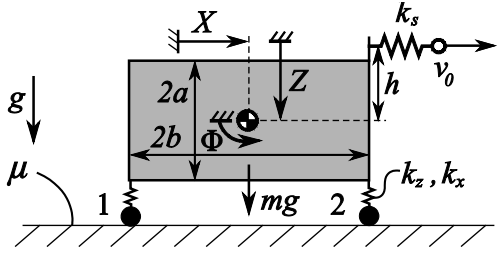

**Fig. 3** Simplified rigid body model with three degrees of freedom. Contact spots are modelled as linear springs

The spots are modelled as linear springs with normal stiffness $k_z$ and tangential stiffness $k_x$ [1]:

$$k_z = E^* D_c \text{ and } k_x = G^* D_c. \quad (9)$$

Here $E^*$ and $G^*$ denote the effective elastic modules that are a function of the shear modulus $G$ and Poisson's ratio $\nu$ of the elastic specimen [1]:

$$E^* = \frac{2G}{1-\nu} \text{ and } G^* = \frac{4G}{2-\nu}. \quad (10)$$

The contact diameter $D_c$ depends on the instantaneous contact configuration and is estimated as shown in Appendix A2. The deflections of the springs in the normal and tangential direction $U_z$ and $U_x$ depend on the actual state of motion of the rigid body. These deflections represent the elastic deformations of the specimen in the normal and tangential direction that are concentrated in the vicinity of the contact spots [32]. Consequently, the contact forces acting on the rigid body are given as:

$$F_{z1/2} = k_z U_{z1/2} \text{ and } F_{x1/2} = k_x U_{x1/2}, \quad (11)$$

where the subscript $(.)_{1/2}$ denotes the two contact spots as shown in Fig. 3. Finally, there remain only three degrees of freedom: $X$ and $Z$, which describe the lateral and the vertical translation of the centre of gravity of the specimen, and $\Phi$, which describes the rotation of the rigid body. In addition, there are two dependent variables for each contact spot: the spring deflections $U_{z1/2}$ and $U_{x1/2}$. Still, the moving base acts as the only external excitation, where $h$ denotes its lever arm with respect to the centre of gravity. This configuration allows the lever arm to be larger than the height $a$.

There is an analogy to the model proposed by Martins et al. which consists of a rigid block that exhibits the same degrees of freedom, i.e. translation and rotation [15]. In the normal direction the rigid block is as well only subjected to its own weight. And in the tangential direction it is restrained by a spring and a viscous damper (dash-pot) and subjected to a moving belt. Also the interface friction is modelled assuming Coulomb dry friction with contact stiffness in the normal and tangential direction. This enables to compare the results. However, their stress distribution in the contact surface is a directly determined function of the rigid body motion and does not enable parts of the contact to slip, while others stick as in the model introduced here. Thus, on the one hand our model is an extension of the model proposed by Martins et al. which also takes into account the spatial variation of stick and slip zones. On the other hand, we neglect the non-linearity of the interface response.

The basic mechanisms that are responsible for the experimentally observed dependency of the frictional resistance on the system dynamics are the same, i.e. vibrations in the normal direction and coupling effects between normal, tangential and rotational degrees of freedom. In our model, a rotation of the body leads to varying normal and tangential forces at both contacts. The other way round, this asymmetry acts as a twisting moment on the rotation. Thus, an appropriate synchronization of these motions will enable the system to walk in such a way as explained in the beginning of this section. The questions remain, under which conditions the excitation of the constantly moving base can cause self-excited oscillations of the rigid body and whether these oscillations will be synchronized in the proposed manner.

### 2.2. Parameters of influence

In order to reduce the number of parameters, all degrees of freedom and the time are replaced by a non-dimensional representation. We normalize with the geometrical constants $a$ and $b$, what yields:

$$x = \frac{X}{b}, \; z = \frac{Z}{b}, \; \varphi = \frac{a}{b}\Phi \quad (12)$$

and:

$$u_{z1/2} = \frac{U_{z1/2}}{b}, \; u_{x1/2} = \frac{U_{x1/2}}{b}. \quad (13)$$

In addition, we introduce the characteristic period $\tau$ and the dimensionless time $t$:

$$\tau = \sqrt{\frac{m}{k_x}} \Rightarrow t = \sqrt{\frac{k_x}{m}} T. \quad (14)$$

This yields the accelerations as:

$$\ddot{X} = \frac{k_x b}{m} x'', \; \ddot{Z} = \frac{k_x b}{m} z'', \; \ddot{\Phi} = \frac{k_x b}{ma} \varphi'', \quad (15)$$

where $(.)'$ denotes the derivation with respect to the non-dimensional time $t$. This procedure reduces the number of parameters from eleven to eight. The remaining ones are listed in Table 1.

**Table 1** definition and interpretation of the non-dimensional parameters of influence

| parameter | definition | physical interpretation |
|---|---|---|
| $\kappa_1$ | $\dfrac{k_s}{k_x}$ | ratio of macroscopic and microscopic stiffness |
| $\kappa_2$ | $\dfrac{k_z}{k_x}$ | ratio of normal and tangential stiffness |
| $\kappa_3$ | $\dfrac{a}{b}$ | geometry ratio, slenderness of the body |
| $\kappa_4$ | $\dfrac{mg}{k_x b}$ | ratio of normal force and contact stiffness |
| $\kappa_5$ | $\mu$ | microscopic friction |
| $\kappa_6$ | $\dfrac{v_0}{b}\sqrt{\dfrac{m}{k_x}}$ | velocity dependent non-dimensional parameter |
| $\kappa_7$ | $\dfrac{\Theta}{mb^2}$ | ratio of rotational inertia of the rigid body |
| $\kappa_8$ | $\dfrac{h}{a}$ | ratio of the lever arm of the excitation |

Considering the analogy to the model of Martins et al. [15] the parameter $\kappa_1$ corresponds to their stiffness parameter $s_M$ that takes into account the ratio of the stiffness of the base spring and the normal contact stiffness. In addition parameter $\kappa_3$ is equivalent to their ratio of height and width $h_M$. Finally, the microscopic friction $\mu$ resembles to their more sophisticated friction parameter $f_M$. Despite the differences between the two models this enables to compare the overall trend of the influences of the parameters.

### 2.3. Simulation

In the non-dimensional form of the system, the spring deflections correspond to the contact forces:

$$f_{N1/2} = \kappa_2 u_{z1/2} \text{ and } f_{x1/2} = u_{1/2}. \tag{16}$$

Assuming small rotations, i.e. $\varphi \ll 1$, Newton's second law initially yields the equations of motion (EoM) for the system as:

$$x_s' = \kappa_6, \tag{17}$$

$$x'' = -u_{x1} - u_{x2} + \kappa_1(x_s - x + \kappa_8\varphi), \tag{18}$$

$$z'' = -\kappa_2 u_{z1} - \kappa_2 u_{z2} + \kappa_4, \tag{19}$$

$$\varphi'' = \frac{1}{\kappa_7}\Big\{-\kappa_2\kappa_3 u_{z1}(1-\varphi) + \kappa_2\kappa_3 u_{z2}(1+\varphi)$$
$$-u_{x1}(\kappa_3^2 + \varphi) - u_{x2}(\kappa_3^2 - \varphi) \tag{20}$$
$$-\kappa_1(\kappa_3^2\kappa_8 + \varphi)(x_s - x + \kappa_8\varphi)\Big\},$$

where $x_s = \kappa_6 t$ denotes the base displacement. We assume the springs to be undeflected for $x, z, \varphi = 0$.

As shown in Fig. 4 the contact spots might be in contact or be released from the substrate depending on the instantaneous motion. Here, the left contact (1) is fully released, what will lead to zero contact forces.

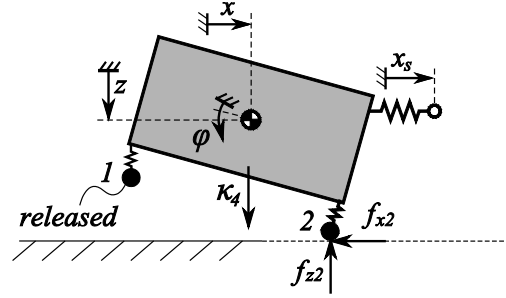

**Fig. 4** Free body diagram of the non-dimensional system. Rotation causes release of left contact from the substrate

In case that a spring is in contact, the spots can slip or stick, depending whether the instantaneous friction bound is exceeded or not. These many options are taken into account using a case distinction. The equations of motion are solved stepwise using the so called Verlet algorithm [33]. In this, we first compute the state of motion of the system $x(t_{n+1})$, $z(t_{n+1})$, $\varphi(t_{n+1})$ in the time step $t_{n+1}$. Afterwards, we compute the normal and tangential deflections, i.e. the contact forces. These determine the velocities in time step $t_{n+1}$ that in turn determine the state of motion in time step $t_{n+2}$ and so on. Initially, the *test* normal deflections are calculated as:

$$\tilde{u}_{z1}(t_{n+1}) = z(t_{n+1}) + \kappa_3^{-1}\varphi(t_{n+1}), \tag{21}$$

$$\tilde{u}_{z2}(t_{n+1}) = z(t_{n+1}) - \kappa_3^{-1}\varphi(t_{n+1}). \tag{22}$$

These are used in the first case distinction which captures released contacts and reads:

**if** $\tilde{u}_{z1/2}(t_{n+1}) > 0$ **then**
$$u_{z1/2}(t_{n+1}) = \tilde{u}_{z1/2}(t_{n+1}), \tag{23}$$
**else**
$$u_{z1/2}(t_{n+1}) = 0. \tag{24}$$
**end**

After that, the tangential deflections are computed using a second case distinction. If a contact sticks, the change of tangential deflection corresponds directly to the change of motion of the system. Thus, the test deflections yield:

$$\tilde{u}_{x1/2}(t_{n+1}) = u_{x1/2}(t_n) + \Delta t x'(t_n) + \Delta t \varphi'(t_n). \tag{25}$$

The contacts slip if these *test* deflections exceed the instantaneous friction bound. Consequently, the second case distinction reads:

**if** $|\tilde{u}_{x1/2}(t_{n+1})| \geq \mu\kappa_2 u_{z1/2}(t_{n+1})$ **then**
$$u_{x1/2}(t_{n+1}) = \mu\kappa_2 u_{z1/2}(t_{n+1}) \cdot \text{sign}(\tilde{u}_{x1/2}(t_{n+1})) \tag{26}$$
**else**
$$u_{x1/2}(t_{n+1}) = \tilde{u}_{x1/2}(t_{n+1}). \tag{27}$$
**end**

In order to analyse the frictional resistance, we determine the motion of the system in steady state. The *mean* of the excitation spring force of the constantly moving base:

$$f_s = \kappa_1 (x_s - x + \kappa_8 \varphi) \tag{28}$$

corresponds to the *mean* resistance force of the system. Finally, with $\langle . \rangle$ being the average over period of observation, the effective coefficient of friction $\mu_e$ is computed as:

$$\mu_e = \frac{\langle f_s \rangle}{\mu \kappa_4}. \tag{29}$$

## 3. Theoretical results

Firstly, using the simulation model we identify appropriate parameter ranges of $\kappa_{1-8}$ that lead to a reduction of the effective friction $\mu_e$. This is followed by a detailed analysis in order to identify the theoretical maximal reduction.

### 3.1. Influence of the parameters

The non-dimensional model is described by eight parameters, each of if potentially influencing the motion and the effective frictional resistance. Starting point for the analysis is the determination of basic points in parameter space.

The ratio of base-spring stiffness and contact stiffness $\kappa_1$ can be interpreted as a measure for the stiffness of the system. Hence, high values correspond to a stiff tribological system as a disc brake. One example for a soft system is a pin-on disc tribometer with flexible arm. We assume the range of $\kappa_1$ to be:

$$\kappa_1 = \frac{k_s}{k_x} \Rightarrow \kappa_1 \in [10^{-1}, 10]. \tag{30}$$

The so called Mindlin ratio $\kappa_2$ depends on the contact geometry. Its bandwidth is narrow:

$$\kappa_2 = \frac{k_z}{k_x} \Rightarrow \kappa_2 \in [1, 1.4]. \tag{31}$$

Parameter $\kappa_3$ denotes the ratio of height $2a$ and width $2b$ of the rigid body, i.e. the slenderness. We consider medium values:

$$\kappa_3 = \frac{a}{b} \Rightarrow \kappa_3 \in [0.5, 1.5]. \tag{32}$$

Parameter $\kappa_4$ lacks an illustrative physical meaning. However, we can estimate it assuming a specimen of mass $m = 100$ g and half-width $b = 1$ cm and a tangential contact stiffness of $k_x \in [10^4, 10^5]$ N/m:

$$\kappa_4 = \frac{mg}{k_x b} \Rightarrow \kappa_4 \in [10^{-4}, 10^{-2}]. \tag{33}$$

In this first step, we assume medium values of the microscopic coefficient of friction:

$$\mu \in [0.2, 0.6]. \tag{34}$$

The parameter $\kappa_6$ depends inter alia on the speed of the constantly moving base $v_0$. Assuming a velocity range of $v_0 \in [1, 100]$ mm/s it results to:

$$\kappa_6 = \frac{v_0}{b} \sqrt{\frac{m}{k_x}} \Rightarrow \kappa_6 \in [10^{-4}, 10^{-2}]. \tag{35}$$

Assuming a rectangular shape of the rigid body and taking into account (32), the ratio of the rotational inertia $\kappa_7$ results to:

$$\kappa_7 = \frac{\Theta}{mb^2} = \frac{4}{12}(1 + \kappa_3^2) \Rightarrow \kappa_7 \in [0.5, 1.5]. \tag{36}$$

It should be emphasized that $\kappa_7$ is independent from $\kappa_3$ as the exact shape of the rigid body can differ from the rectangular shape. Finally, $\kappa_8$ denotes the ratio of the lever arm and the height. Allowing the spring to be applied underneath the centre of gravity, the range of $\kappa_8$ results to:

$$\kappa_8 = \frac{h}{a} \Rightarrow \kappa_8 \in [-0.8, 0.8]. \tag{37}$$

In order to determine the influence of the parameters we make use of the so called design of experiments (DoE) approach. This method systematically minimizes the effort needed for a sufficiently accurate analysis. We use a so-called full-factorial experimental schedule that considers all possible basic points in parameter space. Thus, we take three values of each parameter (minimum, intermediate, maximum) and compute the motion of the system in steady state to give the effective coefficient of friction $\mu_e$. Overall, this leads to $3^8 = 6561$ different combinations. In order to enhance the overall computation time, the calculation is performed in parallel on a graphics processing unit (GPU). Using a Geforce GTX 560 GPU and a step size of $\Delta t = 10^{-3}$ the whole calculation takes about 15 min for a non-dimensional period of observation of 4000. We calculate the main effect, which is the mean of all $6561/3 = 2187$ combinations in which one specific parameter is held constant.

Fig. 5 depicts the main effect on the effective coefficient of friction $\mu_e$. The overall influence of factors is rather low. The maximal reduction being in the range of 10%. On the one hand, this may mean that the individual influence of all parameters is low and there do not exist any points in the parameter space that lead to a strong reduction. On the other hand, this may indicate that only a few parameters are from great influence, whereas the majority has almost no influence. Thus, a strong reduction would only occur in a few points in parameter space. In this work the DoE analysis rather serves as a tool for the identification of convenient parameter combinations. Despite the fact that a further analysis is given later in section 3.2 and section 3.3 we give a short interpretation of the results. However, we want to emphasize that the results of the DoE should be considered with caution, as the overall effect is rather low.

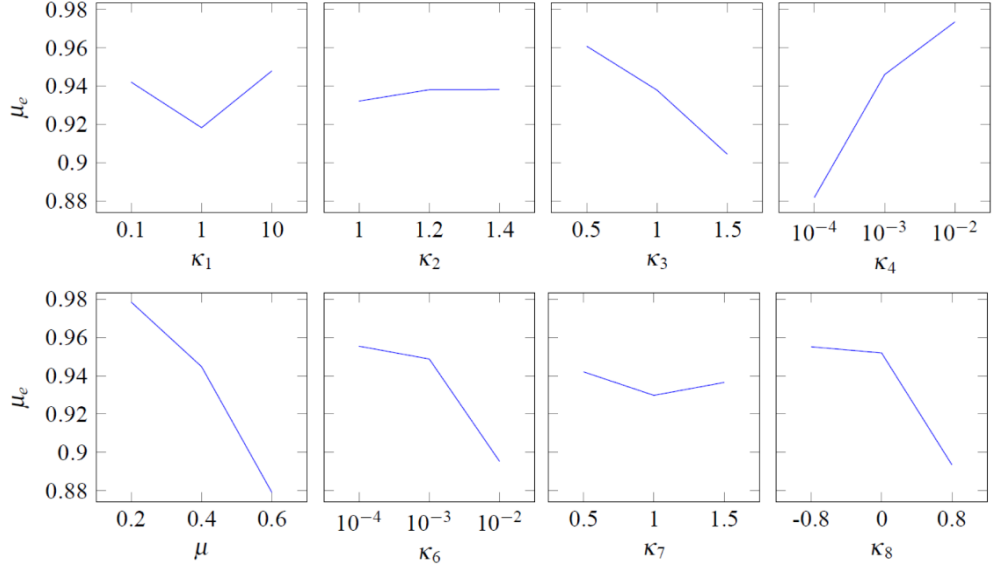

**Fig. 5** Main effect-plot of the eight parameters of influence on the effective coefficient of friction

More specifically, it turns out that $\mu_e$ is low for $\kappa_1$ around 1, meaning that a similarity of the excitation and contact stiffness is convenient. This corresponds to the results of Martin et al. according to which a large stiffness parameter $s_M$ leads to an apparently smooth sliding with an effective coefficient of friction lower than the static one [15]. As one expects the narrow range of the Mindlin-ratio $\kappa_2$ results in a low influence. In contrast, $\kappa_3$ is from great influence indicating that a more compact shape of the rigid body decreases $\mu_e$. A higher $\kappa_3$ leads to a higher rotational moment of the spring force $f_s$ in comparison to the ones of the contact forces. Conversely, the friction increases with increasing $\kappa_4$. Thus, low ratio of mass and contact stiffness reduces the effective friction of this system. The influence of the microscopic friction $\mu$ is relatively high, where $\mu_e$ decreases with increasing $\mu$. This effect is consistent with the beam model by Adams [25] and can be explained by an increasing interplay of the tangential forces and the rotational moment that leads to self-excited oscillations. The friction also decreases for increasing $\kappa_6$ indicating that the effective friction decreases with increasing velocity. This effect also occurs in the slip wave model of Adams [28] and in the rigid body model of Martins et al. [15]. In addition, the effect of decreasing friction with increasing velocity was experienced in numerous experiments [9-12] where an overview is given for example in [13]. The influence of $\kappa_7$ is relatively poor. In contrast, the lever arm of the base, which is represented by $\kappa_8$, is from great influence. The effective friction decreases with increasing lever arm, thus with an increasing moment of the spring force $f_s$ with respect to the contact spots. This increases the coupling between the rotational moment and the friction forces and is consistent with the beam model [25].

In addition, we compute the main effect on the minimal vertical amplitude in steady state $z_{min}/z_{stat}$. Here $z_{stat}$ is the static displacement that is given as:

$$z_{stat} = \frac{\kappa_4}{2\kappa_2}. \qquad (38)$$

A negative $z_{min}$ indicates jumping of the rigid body, i.e. a total release of the contact spots. A high magnitude of $z_{min}$ indicates instabilities and self-excited oscillations of the rigid body in the normal direction that are caused by the interplay of the friction force and the rotational moment. It shows that $\kappa_2$, $\kappa_3$ and $\kappa_7$ all have a weak effect on the vertical amplitude. The absolute value increases with decreasing $\kappa_1$, what shows that the vertical motion is stronger, the weaker the guidance of the spring. Decreasing $\kappa_4$ strongly increases the magnitude of $z_{min}$. Thus, a heavier specimen jumps less strongly for the same contact stiffness and width $b$. The magnitudes of $z_{min}$ strongly increase with $\mu$. This may be explained with the effect that a higher friction increases the coupling between lateral and vertical translation and rotation, what leads to higher displacements in the vertical direction. This is again consistent with the Adams beam model where increasing $\mu$ leads to increasing instabilities [25]. The same effect occurs in the Martins et al. model, where the normal displacement increases with the friction parameter $f_M$ [15]. Parameter $\kappa_6$ has a strong effect, indicating that a higher velocity generates higher vertical jumps. One can assume that a higher velocity simply increases the kinetic energy of the system that is then transferred to the vertical motion through the coupling effect. In addition, it shows that the vertical amplitudes are weak for $\kappa_8 = 0$.

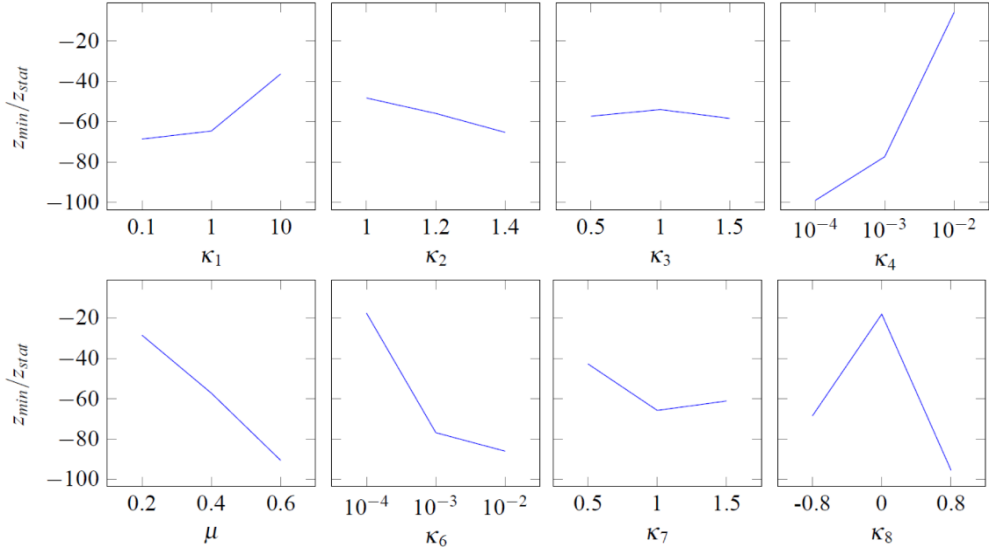

**Fig. 6** Main effect-plot of the eight parameters of influence on the vertical minimal amplitude

Taken together, the DoE analysis shows that a relatively narrow parameter range needs to be matched in order to induce a reduction. Especially for the parameters $\kappa_4$, $\mu_0$ and $\kappa_6$ the reduction is accompanied with an increasing amplitude of the oscillation in the vertical direction. Again it should be emphasized that the DoE analysis is used to determine parameter combinations that enable a significant reduction. It is not the intention to give a detailed analysis at this point. However, the results are consistent with those found in the literature.

### 3.2. Reduction of the frictional resistance

We keep some of the parameters constant in order to give a more detailed analysis. The results of section 3.1 show that a significant reduction requires $\kappa_1$ around 1 and low values of $\kappa_4$. In addition, the influence of $\kappa_2$ and $\kappa_7$ is relatively poor and high values of $\kappa_6$ lead to a strong amplification of the vertical amplitudes. We keep these and the parameter $\kappa_3$ fixed, where the exact values are listed in Table 2. A low value for $\kappa_6$ is chosen in order to induce only small vertical amplitudes in the range of the static displacement $z_{stat}$. This maintains a microscopic character of the oscillations. The remaining parameters are the microscopic friction $\mu$ and the position of the lever arm $\kappa_8$. According to the DoE analysis, both have a significant influence on the effective friction. In addition $\kappa_8$ is relatively easy to vary in real experiments and the influence of $\mu$ was already examined by other authors [25, 15] what gives the possibility for comparison.

**Table 2:** Parameters that are constant in the detailed study

| parameter | $\kappa_1$ | $\kappa_2$ | $\kappa_3$ | $\kappa_4$ | $\kappa_6$ | $\kappa_7$ |
|---|---|---|---|---|---|---|
| value | 2 | 1.2 | 0.5 | $10^{-4}$ | $10^{-4}$ | 0.42 |

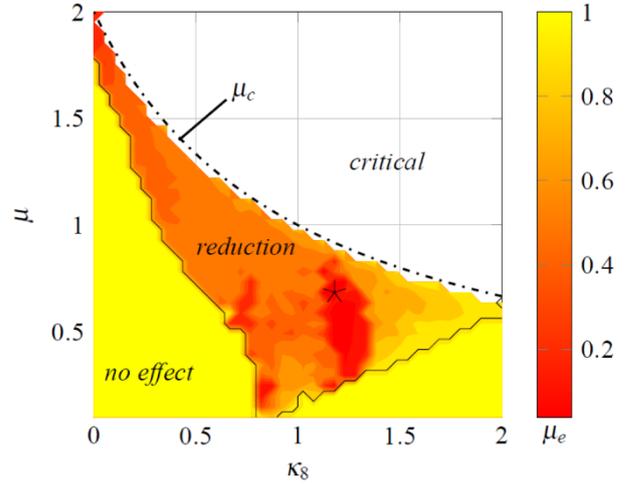

**Fig. 7** Parameter study of $\kappa_8$ and $\mu$ for the effective coefficient of friction $\mu_e$. The star denotes the maximal reduction of 98%. The dash-dot line gives $\mu_c$ of (39)

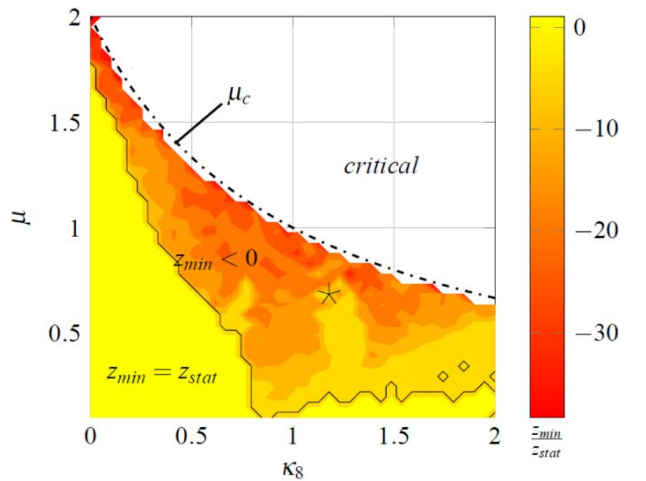

**Fig. 8** Parameter study of $\kappa_8$ and $\mu$ for the minimal vertical amplitude $z_{min}/z_{stat}$. The star denotes the maximal reduction of 98%. The dash-dot line gives $\mu_c$ of (39)

We conduct a parameter study for 1600 combinations of the two remaining factors. Again, we examine the influence on the effective friction $\mu_e$, which is shown in Fig. 7, and the influence on the minimal vertical amplitude, which is shown in Fig. 8. In both figures occur three characteristic parameter ranges, namely the *no effect-range*, the *reduction range* and the *critical range*. These can be explained as follows.

*No effect range:* Here no reduction of the macroscopic frictional resistance occurs at all, i.e. $\mu_e \approx 1$. This range is indicated by the solid line in Fig. 7. The comparison of Fig. 7 and Fig. 8 shows that this parameter range coincides with a region of low vertical amplitudes or even of no vertical vibrations at all. The region of constant vertical displacement is delimited by the solid line in Fig. 8. This indicates that vertical vibrations play an important role in the reduction effect. Without vertical vibration, no reduction occurs at all. This is consistent with the known experiments of Tolstoi et al. [17, 18].

*Critical range:* The reason of the critical behaviour of the system is caused by a tilting of the rigid body over the right spot. It is indicated by the white regions in Fig. 7 and Fig. 8. The simulation is stopped in this case, because the model as stated in (17)-(20) is no longer valid as the magnitudes of $\varphi$ no longer permit linearization of the trigonometric functions. Using the principle of angular momentum with respect to contact 2 and the condition that tilting in the clockwise direction requires a negative angular acceleration gives the critical coefficient $\mu_c$ for the limiting case:

$$\mu_c = \frac{1}{\kappa_3 (1+\kappa_8)} \ . \quad (39)$$

Relation (39) is denoted by the dash-dot line in Fig. 7 and Fig. 8 and is in good agreement with the simulation results. Within the critical limit, the rotational amplitudes where small, i.e. $\varphi < 10^{-3}$. A similar type of critical system behaviour was also studied by Martins et al. for their sliding block model. In this, the maximum value $\bar{f}_M$ of the friction parameter $f_M$ was chosen such that steady sliding equilibrium ceases to be possible due to tumbling of the block [15]. The case $\kappa_3 = 1$ and $\kappa_8 = 1$ exactly corresponds to the case $\bar{f}_M = 1/h_M$, where $h_M$ is the ratio of height and length of the block. They conclude that no one would run a comparable experiment allowing for the occurrence of such large oscillations. However, they point out that the same may not be true for instance in a pin-on-disk tribometer having very flexible arms and a small contact region [15].

*Reduction range:* For a specific range of parameters occurs a significant reduction of $\mu_e$ over 50%. The comparison of Fig. 7 and Fig. 8 shows that the reduction coincides with negative vertical amplitudes $z_{min}$. This indicates that a total release of the contact spots, i.e. jumping, is an important prerequisite for a reduction. The maximal reduction of 98% is denoted by the star and occurs for $\kappa_8 = 1.18$ and $\mu = 0.68$. The corresponding minimum vertical amplitude is $z_{min}/z_{stat} = -0.42$. We conclude that one reason for the reduction are oscillations of the rigid body in the vertical direction. This is consistent with the well-known experiments of Tolstoi et al. [17, 18] and with theoretical works [15, 19, 20, 25]. However, the amplitudes are of the order of the static vertical displacement. Thus, in a real system these vibrations are from a microscopic character and can be superposed unnoticed to an apparently smooth sliding motion.

For a further analysis, we compute the relative stick time of the contact spots, which is the ratio of the time a specific contact sticks $T_{s1/2}$ and the period of observation $T$:

$$t_{s1/2} = \frac{T_{s1/2}}{T} \ . \quad (40)$$

The stick times are depicted in Fig. 9 and Fig. 10. It shows that sticking coincides with the reduction effect with $t_{s1} \approx 15\%$ and $t_{s2} \approx 20\%$ in the reduction range. In contrast, no sticking occurs in the no effect range.

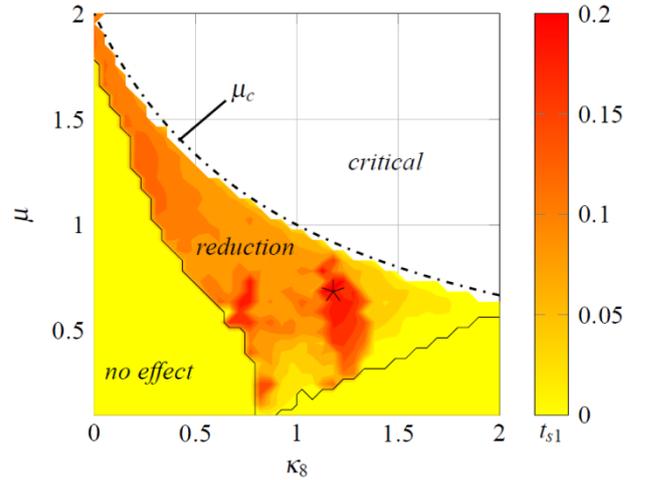

**Fig. 9** Parameter study of $\kappa_8$ and $\mu$ for the relative stick time $t_{s1}$ of contact 1. In the reduction range applies $t_{s1} \approx 15\%$

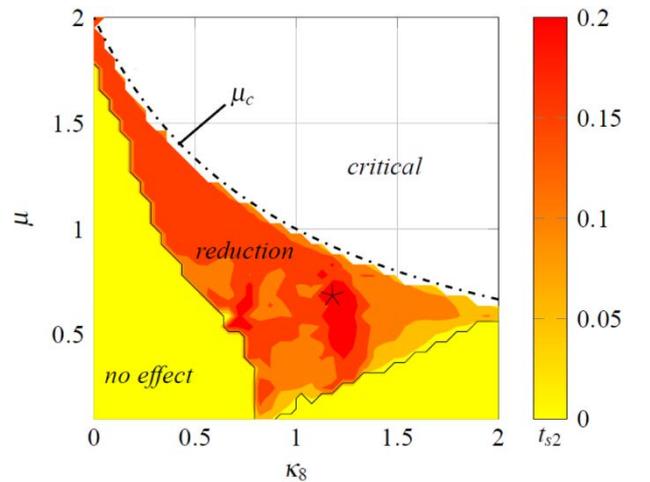

**Fig. 10** Parameter study of $\kappa_8$ and $\mu$ for the relative stick time $t_{s2}$ of contact 2. In the reduction range applies $t_{s2} \approx 20\%$

There are three explanations for this. Firstly, sticking leads to storage of elastic energy in the contacts. This energy is needed for the jumping that occurs in the reduction range, as shown in Fig. 8. Secondly, it is possible that only one spot is sticking, while the other one is slipping or is completely released from the substrate as introduced in section 2. And thirdly, sticking may decrease the average value of the friction force as observed by Martins et al. [15].

Furthermore, we compute the Pearson correlation coefficients $r_1$ and $r_2$ [34] to examine the linear correlation between the velocity of a spot $s'_{1/2}$ and the corresponding normal spring deflection $u_{z1/2}$, i.e. the corresponding normal force.

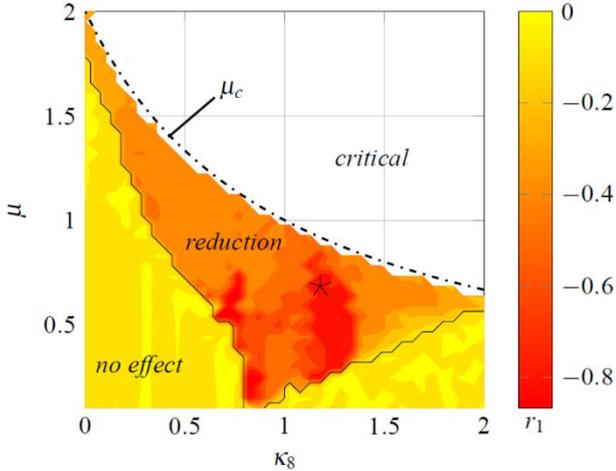

**Fig. 11** Parameter study of $\kappa_8$ and $\mu$ for the correlation coefficient $r_1$ of velocity $s'_1$ and normal deflection $u_{z1}$ of contact 1. In the reduction range applies $r_1 < -0.5$

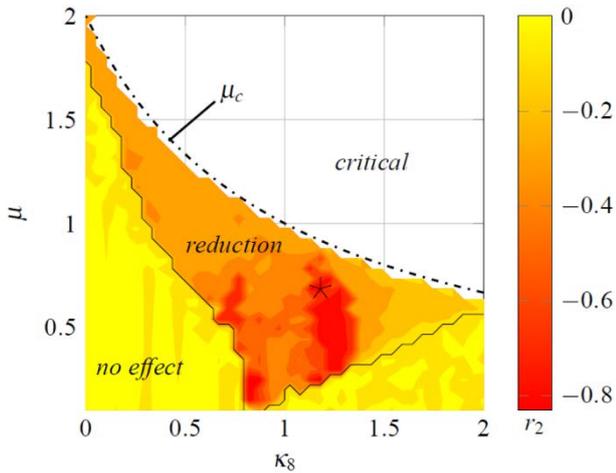

**Fig. 12** Parameter study of $\kappa_8$ and $\mu$ for the correlation coefficient $r_2$ of velocity $s'_2$ and normal deflection $u_{z2}$ of contact 2. In the reduction range applies $r_2 < -0.5$

The velocities are computed as the central difference of the motion of the spots divided by the time step $\Delta t$:

$$s'_{1/2}(t) = \frac{1}{2\Delta t}\left(s_{1/2}(t+\Delta t) - s_{1/2}(t-\Delta t)\right), \quad (41)$$

where the motion of spots $s_{1/2}$ is given as:

$$s_{1/2}(t) = x(t) + \varphi(t) - u_{x1/2}(t). \quad (42)$$

Fig. 11 and Fig. 12 give the correlations for the spots. The case $r = 1$ describes a positive linear correlation between two quantities. A comparison of the different parameter ranges in both figures shows that the correlations are about -0.5 in the reduction range. This indicates a negative linear correlation of spot velocities and normal forces as low forces coincide with high velocities and vice versa. This means that the spots make most of their motion while the corresponding normal forces are low. In addition, they stick, i.e. $s_{1/2} = 0$, while the corresponding normal force is high. Firstly, this indicates that the specimen walks as proposed in section 2. Secondly, this indicates that the rigid body and its contact spots make most of the forward motion while being totally released from the substrate. This is in particular confirmed by comparison of Fig. 7 and Fig. 8, what shows that the reduction coincides with negative vertical amplitudes $z_{min}$, i.e. jumping of the rigid body.

### 3.3. Vanishing frictional resistance

In order to illustrate the proposed effects, we consider the maximal reduction case that is symbolized by the black star in Fig. 7 – Fig. 12. The corresponding parameter combination is given as:

$$\kappa_{1-8} = \left(2, 1.2, 0.5, 10^{-4}, 0.68, 10^{-4}, 0.42, 1.18\right). \quad (43)$$

The maximum reduction is 98% and the corresponding minimal amplitude is $z_{min}/z_{stat} = -0.42$. This indicates jumping of the rigid body, i.e. a complete release of the contact spots. Fig. 13 depicts the phase-space diagram for the relative displacement $x - x_s$ for the maximal reduction combination. The black dot depicts the initial conditions $x(t = 0) = 0$ and $x'(t = 0) = 0$. It shows that after the blue coloured transient process, the system reaches a stable limit cycle (lc) which is dark coloured.

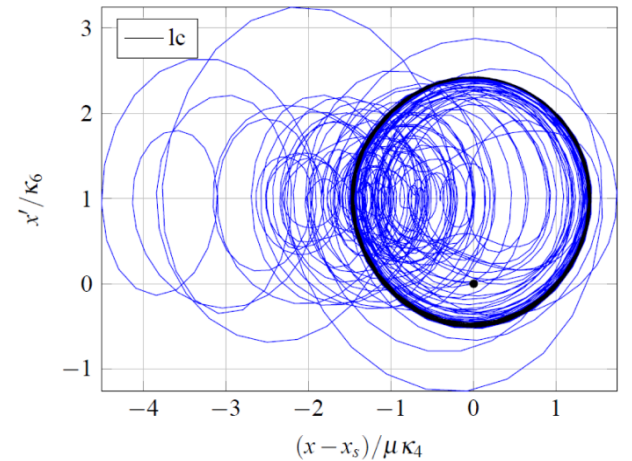

**Fig. 13** Phase space diagram of the lateral motion for the maximal reduction range. The system reaches a stable limit cycle (lc)

Several dynamic quantities in steady state are displayed in Fig. 14, where the difference $\Delta x = x(t) - x(t=0)$ is shown to give all quantities in one plot. The centre of gravity exhibits a harmonic oscillation around the static displacement $z_{stat}$, what leads to highly varying normal deflections of the springs. As one expects, their maximum approximately coincides with the maximum vertical displacement. The small shift between $u_{z1}$ and $u_{z2}$ is caused by the rotation $\varphi$. The vertical motion is highly synchronized with the lateral translation $x$. More specifically, the rigid body makes most of the lateral displacement when the vertical displacement is negative, i.e. the body is jumping, and the contacts are released from the substrate.

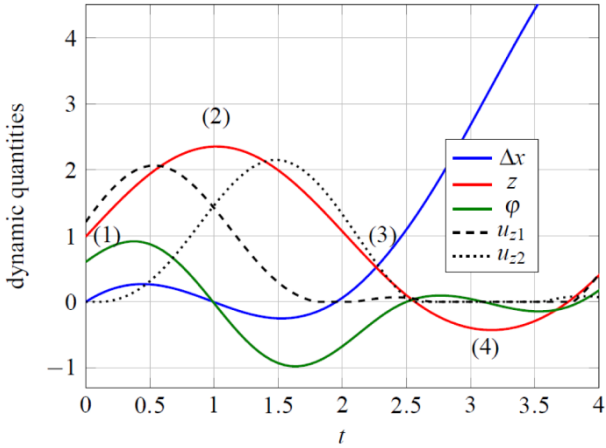

**Fig. 14** Motion of the system for the maximal reduction range. All dynamic quantities are normalized as $\Delta x/\mu\kappa_4$, $z/z_{stat}$, $\varphi/\kappa_3\kappa_4\mu$ and $u_{z1/2}/\kappa_4\mu$

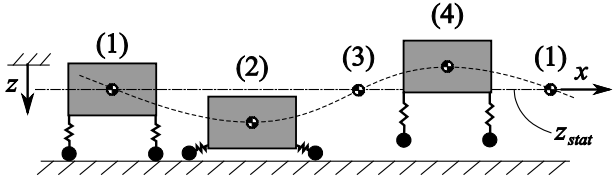

**Fig. 15** Sketch of the motion of the specimen for the maximal reduction parameter combination. The centre of gravity exhibits a harmonic oscillation around the static displacement

We identify four specific states as marked in Fig. 14 and Fig. 15:

(1) start of sticking phase: sticking contact spots, downward movement and increasing normal forces, only minor lateral motion

(2) turning point: end of downward movement, sticking contact spots, high normal forces, storage of elastic energy

(3) end of sticking phase: upward movement of rigid body, i.e. $z < 0$, and start of forward movement

(4) microscopic jump: release of the spots and zero contact forces, fast forward motion of the specimen

The motion can be characterized as a highly synchronized vertical and lateral vibration with alternating jumping and sticking phases of the rigid body. It is this particular system behaviour that leads to an almost vanishing frictional resistance. This matches the effect of micro-walking as introduced in section 2.

### 3.4. Summary

Considering the overall results of section 3, we can finally identify three main effects that are responsible for the reduction effect of the frictional resistance:

1. The coupling of the tangential, vertical and rotational degrees of freedom causes self-excited oscillations in the vertical direction. As the friction forces in the contact spots are mainly responsible for this, the amplitudes grow with increasing microscopic friction and lever arm of the base spring with respect to the contact spots. This effect is consistent with theoretical works of Martins et al. [15] and Adams [25].

2. The experimental works of Tolstoi et al. [17, 18] showed that the normal vibrations themselves already cause a reduction due to the non-linearity between normal separation and normal force. However, in our model a linear dependency was assumed. Hence, the reduction is caused by another effect, namely a strong correlation between low or even zero contact forces and forward motion and high normal forces and sticking contacts.

3. Jumping and fast forward motion of the rigid body requires energy that is stored in the elastic springs. This is maintained by a characteristic alternation between storage and motion phase: the spots stick and energy is conserved that is than used for the vertical jumps and the tangential fast forward motion. Furthermore, sticking simply decreases the average value of the friction force as observed by Martins et al. [15].

We conclude that the dynamic mode responsible for the reduction is the proposed micro-walking effect, i.e. a convenient synchronization of lateral and vertical motion and a jumping of the rigid body with released contacts.

### 4. Experiment

The analysis in section 3 gives clear but also strict guidelines for the design of a real technical system with the effect of frictional resistance reduction through self-excited oscillations. However, there exist certain experimental limits. Firstly, there occur differences in the geometry and the contact configuration between the theoretical and the real system that is described in section 4.1. The differences are mainly caused by the fact that the two-dimensionality of the theoretical model cannot be simply transferred to the real world in particular due to alignment problems. Secondly, there exist several practical limitations which make it impossible to reach the values of the non-dimensional parameters that are required for the almost vanishing friction. For instance, simultaneously fulfilling the conditions $\kappa_1 \approx 1$ and $\kappa_4 \approx 10^{-4}$ is very hard in practice as discussed in Appendix A 3. Thirdly, the results show that there exists a critical coefficient of friction $\mu_c$ that depends on the geometry, respectively the point of application of the spring force as shown in section 3.2. This refers to (39):

$$\mu_c = \frac{1}{\kappa_3 (1+\kappa_8)} . \qquad (44)$$

Thus, the lower the value of $\kappa_8$, the higher the critical coefficient. To avoid the critical range, we limit the experiments to negative values of $\kappa_8$, i.e. the point of application of $f_s$ is always underneath the centre of gravity. This will enhance the stability of the experiment with respect to the rotation, i.e. will avoid tumbling of the rigid body, but will also decrease the reduction. Finally, some of the parameters are easier to vary in the experiments than others. For instance, it does not require much effort to change $\kappa_8$ or the velocity of the base $v_0$. In contrast, changing $\kappa_3$ or $\kappa_7$ requires an update of the geometry and changing $\mu$ requires a different combination of materials. This would lead to undesired changes in several other parameters as $\kappa_1$, $\kappa_4$ and $\kappa_6$ as they depend on material properties. Overall, we consider five parameter sets named set 1-5, whose exact parameters are given in Table 3.

### 4.1. Experimental setting

The experimental set-up shown in Fig. 16 is used to investigate the micro-walking effect. Most important components are the specimen and the substrate. The substrate is mounted on a massive steel plate which is mounted on a linear guide. The drive moves the substrate with velocity $v_0$ relative to the specimen. Thus, the system corresponds to the well-known moving belt model.

The experimental rig is powered by an electro-mechanical testing actuator made by Zwick GmbH & Co. KG whose good ganging properties reduce unwanted dynamic interactions. The actuator is controlled using the controller BX-1. Straight rods act as the spring between specimen and base. The spring force $f_s$ is measured at the base using a S-sensor. We use different sensors, namely the sensor TD-112 by Yuyao Tongda Scales Co., Ltd. and the sensor KD24s by ME measurement systems GmbH. The latter one has smaller dimensions and allows a larger adjustment range of the height of the base. For amplification of the sensor signal, the amplifier system MGC of the company HBM is used, which essentially consists of the amplifier HBM MC 55. The measured signals are fed to the PC via a measuring system with 16-bit data acquisition board, namely NI PCI-6221 by National Instruments Corporation, which also serves to control the drive. The sampling frequency used to measure the spring force is chosen to be $f_{samp} = 4000$ Hz. This satisfies the Whittaker–Nyquist–Kotelnikov–Shannon theorem [35] as:

$$f_{samp} = 4000 \text{ Hz} > 2 \times \Omega_{exp} \approx 760 \text{ Hz} . \qquad (45)$$

Here $\Omega_{exp}$ denotes the characteristic frequency of the spring force $f_s$ which is determined from the frequency spectrum of the time response of $f_s$ for set 5.

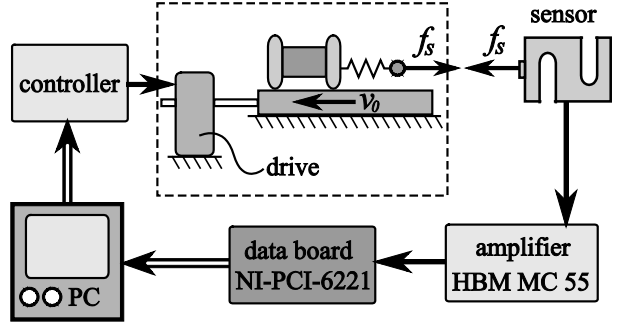

**Fig. 16** experimental rig for the micro-walking effect with measurement chain consisting of force sensor, amplifier, data board and PC

The materials and the exact dimensions of the specimen and the substrate are chosen such that the dimensionless parameter combinations, which are determined with the numerical model as described in section 3.1 and section 3.2., are met as closely as possible. However, experimental feasibility leads to certain limitations. The specimen has an hourglass shape as shown in Fig. 17 (a). It consists of a steel centrepiece and two contact discs made of polypropylene (PP) with Young's modulus $E_{PP} = 1.4$ GPa. This configuration enables a quick adjustment of the geometry and thus inertia properties. The substrate consists of smooth window glass (GS) with Young's modulus $E_{GS} = 90$ GPa and mean roughness $R_a < 10$ nm [36]. It has a prism shape as shown in Fig. 17 (b) and Fig. 18. On the one hand, this configuration enhances the stability and parallelism of the motion of the specimen. On the other hand, the contact configuration of the experiment differs slightly from the initial theoretical model introduced in section 2.1. Thus, for the comparison in section 4.2 we use an extended model as described in Appendix A1. Other dimensions and parameters of the experimental rig and the different combinations set 1-5 are given in Table 3.

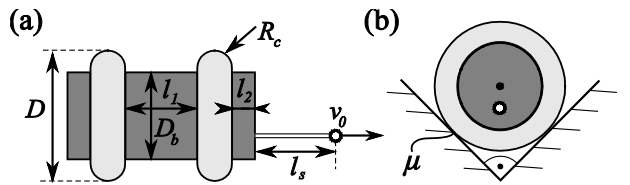

**Fig. 17** Side view of the specimen with dimensions (**a**). Front view of specimen with prism-shaped substrate (**b**)

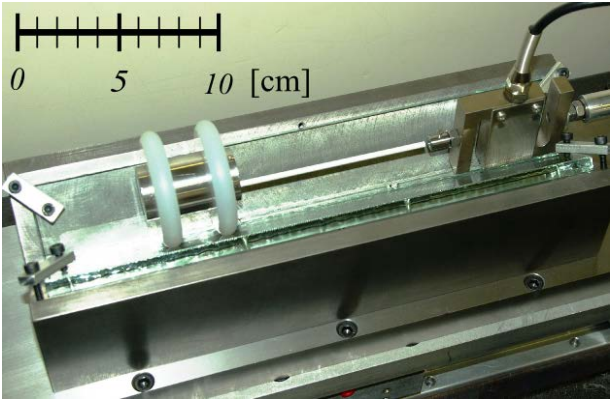

**Fig. 18** Experimental setting. Specimen, prism-shaped glass substrate and s-force sensor

**Table 3** Parameters of the micro-walking experimental rig for the different parameter combinations set 1-5. The numbers in brackets give the corresponding set

| parameter | sym | value |
|---|---|---|
| contact disc radius | $R_c$ | 5 mm |
| contact disc diameter | $D$ | 60 mm |
| body diameter | $D_b$ | 30(1,2,3,4)/50(5) mm |
| model half-height | $a$ | 21.2 mm |
| centre part-length | $l_1$ | 18(1,3,5)/40(2,4) mm |
| model half-width | $b$ | 14(1,3,5)/25(2,4) mm |
| contact disc length | $l_2$ | 5 mm |
| specimen mass | $m$ | 201(1,3)/327(2,4)/490(5) g |
| spring-rod length | $l_s$ | 100 mm |
| friction coefficient | $\mu$ | 0.79(1,2)/0.68(3,4)/0.75(5) |
| Young's modules | $E$ | 1.4 (PP)/90 (GS) GPa |
| Poisson-ratio | $\nu$ | 0.4 (PP)/0.3 (GS) |
| contact diameter | $D_c$ | 1.8 mm |

In order to increase the reproducibility, we briefly describe how the experiments were conducted. The final experimental results are given in section 4.2 and section 4.3. The experimental procedure used for detection of the time responses of the spring force, as for instance shown in Fig. 20, is as follows:

1. alignment of base and specimen
2. measuring run with desired velocity $v_0$
3. return run with velocity $v_0 = 15$ mm/s

The running distance was 190 mm. In order to ensure an almost stationary contact configuration, i.e. elliptical wear patches at the contact zones as described in Appendix A2, we firstly conducted 100 runs for the running-in process after every change of the experimental set-up. In order to study a particular combination of parameters 10 to 20 iterations, i.e. measuring runs, are performed before the experimental set-up was changed.

### 4.2. Experimental results

Due to the reasons stated above, i.e. the effort for a variation of certain parameters, we first compare significantly different parameter ranges without intermediate combinations as listed in Table 4. In order to identify convenient combinations, that are experimentally reachable, the extended model as introduced in Appendix A1 was used. It shows that within the experimental limits, the theoretical maximal reduction was about 50%. However, we initially underestimated the contact stiffness. On this basis an insufficient spring stiffness was chosen such that the stiffness parameter was $\kappa_1 \approx 0.1$ instead of the intended $\kappa_1 \approx 2$. Set 1 meets the theoretical maximal range within the experimental limits except for $\kappa_1$. Set 2 has unsuitable geometry and inertia properties, i.e. $\kappa_3$ is inconveniently changed. Set 3 and set 4 correspond to the previous ones with low $\kappa_1$, i.e. a very soft base-spring. For this purpose, a spiral spring with stiffness $k_s = 60$ N/m was used.

**Table 4** Parameters of the different sets used in the experiment. Parameters $\kappa_2$ and $\kappa_8$ are the same for all

| set | 1 | 2 | 3 | 4 |
|---|---|---|---|---|
| $\kappa_1$ | $\approx 0.1$ | $\approx 0.1$ | $\approx 10^{-4}$ | $\approx 10^{-4}$ |
| $\kappa_2$ | 1.33 | 1.33 | 1.33 | 1.33 |
| $\kappa_3$ | 1.51 | 0.85 | 1.51 | 0.85 |
| $\kappa_4$ | $2 \cdot 10^{-4}$ | $1 \cdot 10^{-4}$ | $2 \cdot 10^{-4}$ | $1 \cdot 10^{-4}$ |
| $\mu$ | 0.79 | 0.79 | 0.68 | 0.68 |
| $\kappa_7$ | 0.95 | 0.25 | 0.95 | 0.25 |
| $\kappa_8$ | -0.47 | -0.47 | -0.47 | -0.47 |

In the experiments, we vary the base velocity $v_0$ that is linear proportional to the non-dimensional parameter $\kappa_6$. We measure the spring force $f_s(t)$ and compute the effective coefficient of friction as in (55). Fig. 19 depicts the results of the different sets. For the sake of clarity, all curves are plotted over $v_0$ and not over $\kappa_6$ as this parameter is cross-influenced by other quantities that change between the sets. The reduction for set 1 is about 54%, meaning that the effective friction $\mu_e$ is more than halved even for $\kappa_1 = 0.1$. In contrast, no significant reduction applies for the other sets. The effective friction slightly varies within a range of approximately 10% of $\mu$. We conclude, that the effect of friction minimization as introduced in section 3 does not occur for sets 2-4, thus $\mu_e \approx 1$. This means, just as with the theoretical results, that in order to induce the friction minimizing effect, a very narrow parameter range has to be met. In addition, the effect hardly depends on velocity but only occurs if a certain velocity limit is exceeded as can be seen for set 1 where $\mu_e$ is halved only for $v_0 > 15$ mm/s. However, the experimental results show that it is possible to induce the effects introduced in section 3.

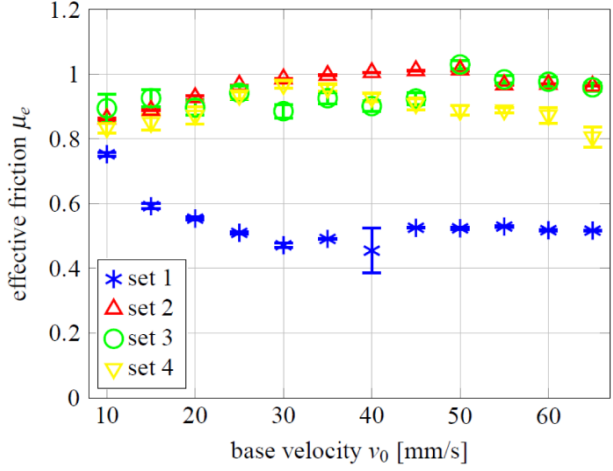

**Fig. 19** Effective coefficient of friction $\mu_e$ for varying base velocity $v_0$ and different parameter sets

In the experiments, the only time varying value that was measured was the spring force in the base. The normal oscillations have not been measured. Nevertheless, we assume that whenever instabilities, i.e. strong oscillations, occur in the normal direction, they will also affect the tangential direction due to the coupling effect. Thus, self-excited oscillations in the normal direction must be detectable in the time response of the base spring. Fig. 20 displays the time response of the spring force $f_s$ for set 1 and set 2. It shows that in case of set 1, which is represented by the blue line, the reduction is accompanied with self-induced oscillations as $f_s$ oscillates with high frequency and magnitude. The force reaches negative values, meaning that the specimen pushes the base point of the spring. For set 2, which is represented by the red line, the force is not oscillating and is positive all the time. We conclude that in this case the instability, i.e. self-excited oscillations in the vertical direction with high amplitude, do not occur at all. Consequently no reduction is induced.

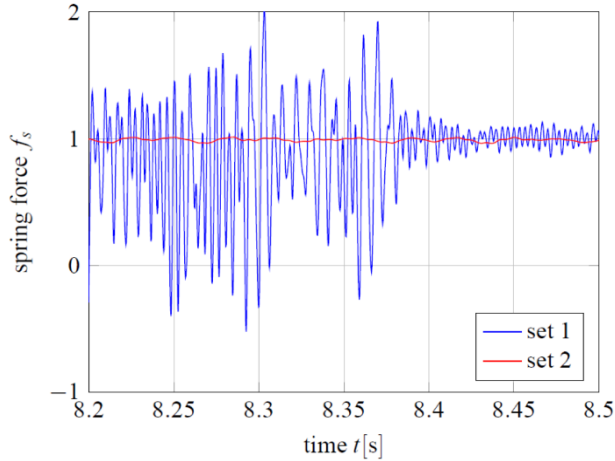

**Fig. 20** Time history of the normalized spring force for set 1 and set 2. Instability occurs only for set 1 but significantly decrease temporarily

Additionally, it shows that the vibrations of the force of set 1 significantly decrease temporarily, as can be seen in Fig. 20 for $t > 8.4$ s. Consequently, the instability as well as the reduction effect temporarily pauses. In order to determine the ratio of the time the instability occurs, i.e. a significant increase of the amplitudes of $f_s$ with negative values, and the time of observation $t_{is}/T$ we compute the revolving variance $var(f_s(t))$ of the friction force. Within the Dirichlet window with size $2t_D$ of the time response, the variance is defined as:

$$var(f_s(t)) = \sqrt{\frac{1}{2t_D/\Delta t} \sum_{t_i=t-t_D}^{t_i=t-t_D} f_s(t_i) - \overline{f}_s(t)} \quad (46)$$

Where $\overline{f}_s(t)$ denotes the corresponding mean within the window:

$$\overline{f}_s(t) = \frac{1}{2t_D/\Delta t} \sum_{t_i=t-t_D}^{t_i=t-t_D} f_s(t_i). \quad (47)$$

Here $f_s(t_i)$ denotes the spring force of the measurement sample $i$ at time $t_i$. The window $2t_D$ corresponds to approximately 10 characteristic periods of $f_s(t)$. As a criterion, we propose that only those time sections are regarded as instable in which $var(f_s(t)) > 0.1$ applies. In addition, for determination of the effective friction $\mu_e$ only instable time sections are taken into account.

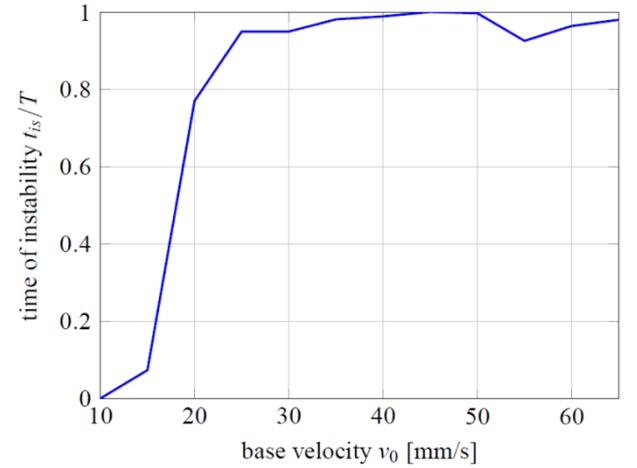

**Fig. 21** Time portion of the instability $t_{is}/T$ as a function of the base velocity $v_0$ for experimental set 1

Fig. 21 gives the relative time of instability $t_i/T$ for set 1. It shows that $t_{is} \approx T$ applies if a certain velocity limit is exceeded. Additionally, the self-excited oscillations highly coincide with minimized friction as can be seen by comparison with Fig. 19. The instability is thus crucial for the reduction effect. This confirms the results of the theoretical model given in section 3.

### 4.3. Comparison of experiment and model

The results of section 3.2 indicate that the application point of the spring is from great influence for the reduction. For this reason, we examine the joint influence of the position of the lever arm, which refers to $\kappa_8$, and the velocity $v_0$, i.e. parameter $\kappa_6$. We consider four values of $\kappa_8$. In the experiment this is simply realized by several differently positioned mounting holes, as shown in Fig. 22 (a). The specimen is rotated such that the line between the mounting hole and the centre of gravity is vertical as in Fig. 22 (b).

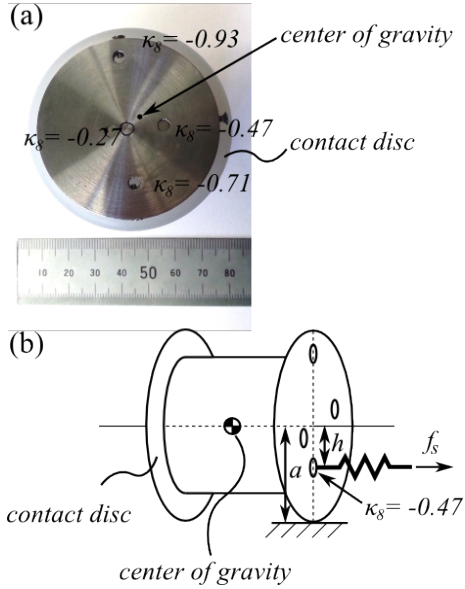

**Fig. 22** Specimen with four positions of the lever arm of the excitation spring with respect to centre of gravity. Four values of $\kappa_8$ (**a**). Lateral view with the line between centre of gravity and mounting hole for $\kappa_8 = -0.47$ being vertical (**b**)

As sketched in Fig. 22 (a) we restrict the experiments to values of $\kappa_8 < 0$ in order to reduce unwanted instabilities of the motion, i.e. tumbling and tilting of the rigid body. This refers to the critical coefficient as defined in (39).

For a mutual verification, the extended contact model was used, that is described in Appendix A1. In comparison to the experiments shown in section 4.2 we used a stiffer material for the connection element and increased the mass of the specimen to lower the influence of unwanted side-effects. With this, the maximal reduction parameter range was identified as listed in Table 5. Theoretically, this combination gives a maximal reduction of about 50%.

**Table 5** Maximum parameter range of the experimental set

| set | $\kappa_1$ | $\kappa_2$ | $\kappa_3$ | $\kappa_4$ | $\mu$ | $\kappa_7$ |
|---|---|---|---|---|---|---|
| 5 | $\approx 2$ | 1.33 | 1.51 | $2 \cdot 10^{-4}$ | 0.75 | 1.77 |

Fig. 23 shows a contour plot of the effective coefficient of friction for 1600 combinations of $v_0$ and $\kappa_8$. Both, higher values of $v_0$ and $\kappa_8$ increase the reduction. Thus, higher rotational moment of spring force with respect to the contact spots leads to higher reduction. Additionally, the yellow area shows the limit range which is delimited by the black line. Here no reduction occurs at all, i.e. $\mu_e \approx 1$. Thus, we conclude that there exists a limit velocity that has to be exceeded in order to induce the reduction effect. And that this limit velocity depends on the lever arm of the spring.

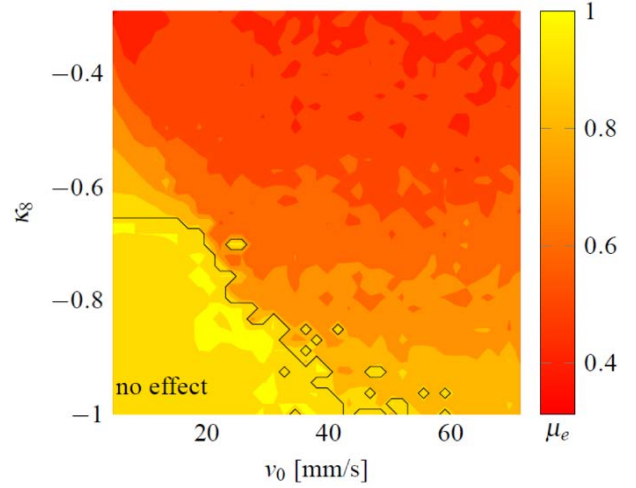

**Fig. 23** Surface plot of the effective coefficient of friction $\mu_e$ as a function of $v_0$ and $\kappa_8$. Computed with the extended model described in Appendix A1

Fig. 24 gives a comparison of the theoretical values, which are symbolized by the circles, and the experimental results, which are denoted by the error-bars and marks. Although there occurs a deviation between theory and experiments, the overall trend is similar. A certain velocity limit exists that has to be exceeded for the reduction effect to occur. For $\kappa_8 = -0.93$ and $\kappa_8 = -0.71$ this velocity threshold is about $v_0 = 40$ mm/s respectively $v_0 = 25$ mm/s. This agrees with the simulation as shown by the corresponding pink and blue circles in Fig. 24. For $\kappa_8 = -0.47$ and $\kappa_8 = -0.27$ the limit velocity is approximately $v_0 = 15$ mm/s. This does not coincide with the simulation, where no limiting velocity exists for these values of $\kappa_8$. However, in the experiments the reduction increases with increasing velocity once the limit velocity is exceeded, what agrees with the simulation. It turns out that for every $\kappa_8$ examined, the specific maximal reduction is higher in the experiments. For $\kappa_8 = -0.93$ the reduction is 52%, for $\kappa_8 = -0.71$ it is 48% and for $\kappa_8 = -0.47$ it is 60% in the experiments. The maximal reduction is 73% and occurs for $\kappa_8 = -0.27$.

In addition, Fig. 25 gives area plots of the time portion of instability $t_{is}/T$ for the experiment for the four cases of $\kappa_8$. The time of instability is determined on basis of the revolving variance as defined in (46). It shows that $t_{is}$ increases with base velocity $v_0$ and $\kappa_8$. Thus, the higher the rotational moment of the spring force of the base with respect to the contacts, the higher is the relative time of instability. In addition, the velocity threshold for the instability to occur is lower for a higher rotational moment. Comparison of Fig. 24 and Fig. 25 shows that the reduction effect also highly coincides with increased time portion of instability.

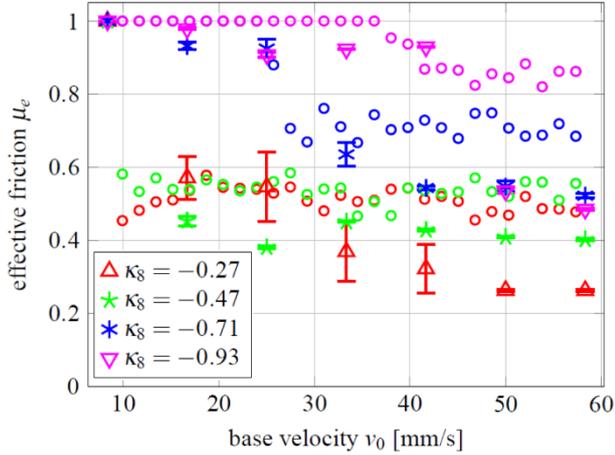

**Fig. 24** effective coefficient of friction $\mu_e$ as a function of base velocity $v_0$ for different $\kappa_8$. Simulation (circles) ad experiment (error bars and marks) show that a velocity limit has to be exceeded to induce the friction minimizing effect

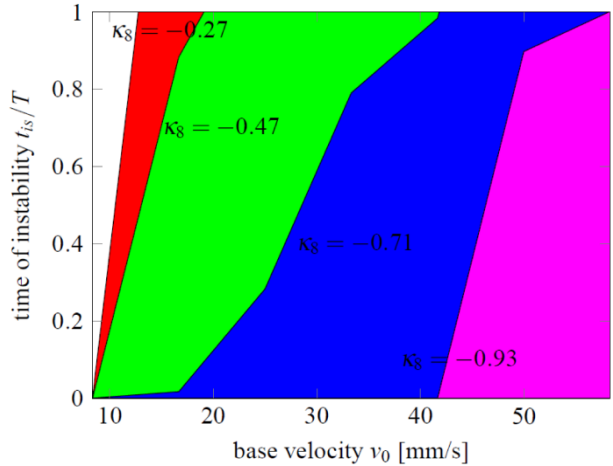

**Fig. 25** Area plot of time portion of instability $t_{is}/T$ for the experiment as a function of base velocity $v_0$ for different $\kappa_8$

The experimental results confirm the results of the DoE as shown in section 3.1 where the reduction increases with increasing $\kappa_8$ and increasing parameter $\kappa_6$, i.e. increasing velocity $v_0$. As stated in section 3.2 the reduction effect is caused by self-excited vertical oscillations of the rigid body. These oscillations increase with an increasing moment of the spring force $f_s$ with respect to the contacts what increases the coupling between the rotational moment and the friction forces. In addition, the oscillations increase with the velocity as these increases the kinetic energy that is transferred to the vertical motion through the coupling effect. The effect of decreasing kinetic friction with increasing velocity was experienced in many experiments [9-12] where an overview is given for example in [13]. The same effect also occurs in theory as in the slip wave model of Adams [28] or in the rigid body model of Martins et al. [15].

However, the deviations between experiment and model are relatively high. One reason for this might be a deviation between the real parameters of the experiment and the corresponding theoretical values. Another one might be that the model does not capture the true contact mechanics and that the influence of the spring on the vertical displacement and the rotation is higher than expected.

## 5. Conclusion

On basis of important experimental and theoretical works a simulation model was introduced, in order to examine the influence of the system dynamics on sliding friction. The model captures known effects as the coupling between normal and tangential motion and adds features as the spatial variation of stick and slip zones. This enables micro-walking of the specimen. By this, we characterize a dynamic mode in which the system travels most of the forward motion while the normal forces in the contacts are low in comparison to the average value. This causes a reduction of the effective coefficient of friction, i.e. a lower macroscopic frictional resistance.

We used numerical simulations and an experimental rig for the analysis. In a first step, we identified parameter combinations for which a significant reduction occurs. The results indicate that a reduction requires that the stiffness of the excitation and the contact stiffness are comparable in size. In addition, the friction decreases with increasing velocity, increasing microscopic friction and increasing coupling between rotational moment and friction forces. These results are consistent with those gained with more or less comparable models [15, 27, 28].

The maximal reduction was 98% in theory and 73% in the experiments. A further analysis of the simulation results shows that the reduction is caused by self-excited oscillations that are induced by the coupling of the different degrees of motion of the micro-walking machine. The excitation causes a rotation of the rigid body which increases the contact forces and induces oscillations in the vertical direction. This instability is characterized by a microscopic jumping of the rigid body with released contacts that is in strong correlation with the lateral motion: low or zero contact forces coincide with a fast forward motion. In addition, we identified the alternation between storage and motion phase as the prerequisite for the characteristic jumping and fast forward movement.

Our model shows that micro-vibrations play an important role for the dynamic influences on the effective frictional resistance of systems that exhibit apparently smooth sliding. In these systems, the experimentally observed dependency on dynamic quantities might be explained by microscopic effects that in fact are influenced by macroscopic system features such as the stiffness or the geometry. One example for this being the rate weakening of the friction coefficient [9-12]. Our model shows that at least in dry contacts this effect can to some extent be explained by dynamic instabilities that simply increase with the sliding velocity. This is particularly important for the design of tribometers. For this purpose the results should be revised in order to give practical guidelines for the design of tribometers. This can be followed by analysis and improvement of an existing tribometer with regard to its dynamic properties.

However, the design-phase of the experimental rig and the experiments itself showed that the parameter range that enables a reduction is narrow and hard to reach. For instance, matching the non-dimensional parameters in reality was very challenging. Furthermore, a change of the scaling of the model would change the non-dimensional parameters, due to cross influences and some constant parameters as the gravitational acceleration. Thus at this stage it remains vague to what extent the proposed effect occurs in practical systems.

**Compliance with ethical standards**

The authors declare that they have no conflict of interest. This article does not contain any studies with human participants or animals performed by any of the authors.

**Appendix**

**A1 Extended contact model**

The contact geometry of the experimental specimen differs slightly from the numerical model. In order to identify appropriate experimental parameter ranges and to enhance comparability, an extended contact model is being used. The prism shaped substrate adds an additional deflection $u_s$ to each contact spring. As sketched in Fig. 26, $u_s$ is aligned perpendicular to the normal and tangential deflections $u_n, u_x$.

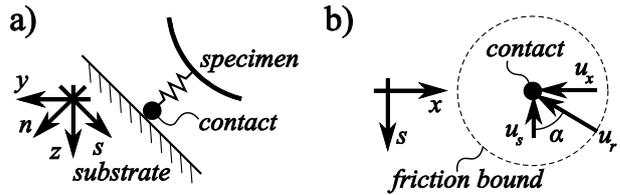

**Fig. 26** Contact spot of extended model with additional spring deflection $u_s$ perpendicular to normal and tangential direction

As mentioned before in case of sticking contacts, the deflections and forces follow directly from the change of motion between time steps. However, in case of slipping contacts, the friction bound:

$$u_{r1/2} = \mu \kappa_2 u_{n1/2} \qquad (48)$$

only gives the norm of deflections $u_x, u_s$. Thus, we have to determine the angle $\alpha$ between $u_x, u_s$ in the new time step. With $\tilde{u}_x, \tilde{u}_x$ being the test deflections, the extended case distinction thus reads:

**if** $\sqrt{\tilde{u}_{x1/2}^2(t_{n+1}) + \tilde{u}_{s1/2}^2(t_{n+1})} \geq u_{r1/2}(t_{n+1})$ **then:**

$$\alpha_{1/2}(t_{n+1}) = \arctan\left(\left|\tilde{u}_{s1/2}(t_{n+1})/\tilde{u}_{x1/2}(t_{n+1})\right|\right), \qquad (49)$$

$$u_{x1/2}(t_{n+1}) = u_{r1/2}(t_{n+1})\cos(\alpha_{1/2}(t_{n+1})), \qquad (50)$$

$$u_{s1/2}(t_{n+1}) = u_{r1/2}(t_{n+1})\sin(\alpha_{1/2}(t_{n+1})), \qquad (51)$$

**else:**

$$u_{x1/2}(t_{n+1}) = \tilde{u}_{x1/2}(t_{n+1}), \qquad (52)$$

$$u_{s1/2}(t_{n+1}) = \tilde{u}_{s1/2}(t_{n+1}), \qquad (53)$$

**end**

where $(.)_{1/2}$ denotes the contact spots on the left and on the right. In fact, there are four contacts spots each exhibiting its own spring deflections. But as the model is symmetrical, the two contacts on the left and on the right obtain the same contact conditions. Thus we consider only two contacts and double the corresponding contact forces:

$$f_{z1/2} = 2 \cdot \kappa_2 \frac{\sqrt{2}}{2}(u_{n1/2}+u_{s1/2}), f_{x1/2} = 2 \cdot (u_{x1/2}). \quad (54)$$

The EoM are the extended model are as in (17) - (20). As a consequence of the inclined contact plane, the effective coefficient must be computed as:

$$\mu_e = \frac{\langle f_s(t) \rangle}{\frac{\sqrt{2}}{2}\mu\kappa_4}. \quad (55)$$

**A2 Contact stiffness and contact diameter**

The contact spots are modelled as linear springs with stiffness:

$$k_z = E^* D_c \text{ and } k_x = G^* D_c. \quad (56)$$

The effective moduli $E^*$ and $G^*$ directly depend on the materials in contact whereas the contact diameter $D_c$ is a function of the actual contact configuration. If we assume the contacts to be of the Hertzian type with an average normal force $\bar{f}_N$ as defined in (5), we get:

$$D_c = \left(\frac{6\bar{f}_N R}{E^*}\right)^{1/3} \approx 0.4 \text{ mm} \quad (57)$$

One drawback of this assumption is that in fact the normal force and thus $D$ change in time. The other drawback is the contact model itself. Due to wear, the shape of the contacts will soon resemble a flattened ellipse with a more or less constant *effective* diameter as shown in Fig. 27. Here one wear spot in the contact zone is shown. A rather crude estimate of the conjugate diameters of the wear spots gives:

$$D_c = 2\sqrt{ab} \approx 2 \text{ mm}. \quad (58)$$

But again, the question remains whether the wear spots really correspond to the contact zones.

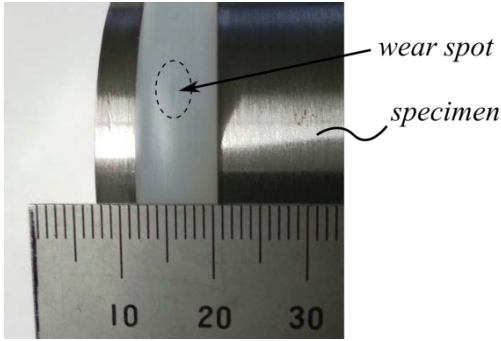

**Fig. 27** Contact zone with wear spot on contact disc. Due to wear the shape soon resembles a flattened ellipse

A third way to estimate the contact stiffness is to compare the characteristic frequencies of the spring forces of the experiment and the simulation. Assuming that the (non-dimensional) simulation maps the reality sufficiently accurate, the characteristic frequencies of experiment and simulation must match, i.e. $\Omega_{exp} = \Omega_{sim}$. In case of the simulation, the characteristic frequency is given as the computed non-dimensional frequency $\omega_{sim}$ divided by the characteristic period $\tau$ of (14). Finally, we can compute the contact stiffness as:

$$k_x = m \left(\frac{\Omega_{exp}}{\omega_{sim}}\right)^2. \quad (59)$$

The characteristic frequencies will be identified using the frequency spectrum of the data set 5 as described in Table 5. The contact stiffness and diameter then yield:

$$k_x = 2 \cdot 10^6 \frac{N}{m} \Rightarrow D_c \approx 1.8 \text{ mm}. \quad (60)$$

Exactly these values are being used for approximation and the comparison of experiment and simulation. However, it should be emphasized that this is a rather crude estimate for the contact stiffness and the real contact mechanics of the experimental system.

**A3 Spring stiffness**

The parameter studies of the numerical model as described in section 3.1 show that the micro-walking effect occurs if for the dimensionless parameters $\kappa_1$ and $\kappa_4$ applies:

$$\kappa_1 = \frac{k_s}{k_x} \approx 1, \kappa_4 = \frac{mg}{k_b b} \approx 10^{-4}. \quad (61)$$

The use of a soft spring, e.g. $k_s = 10^3 \text{ N/m}$, would give experimental advantages as this increases the difference between the frequencies of the specimen and those of the experimental rig, i.e. drive and base frame. However, in order to fulfil (61) the length $b$ would result in completely impractical values, as can be shown for typical dimensions of the experiment, e.g. $m = 100 \text{ g}$, $g = 9.81 \text{ m/s}^2$ g:

$$k_s = k_x \Rightarrow b = 10^4 \frac{mg}{k_s} \approx 10^4 \frac{1 \text{ N}}{10^3 \text{ N/m}} = 10 \text{ m}. \quad (62)$$

Instead, we use stiff polypropylene for the contact discs, such that the tangential stiffness results to $k_x = 2 \cdot 10^6 \text{ N/m}$ and the length to $b = 10^{-2} \text{ m}$. In order to achieve a spring stiffness of $k_s \approx k_x$, we use straight rods made of fibre-reinforced plastic respectively polystyrene. Their corresponding Young's modules were measured to be:

$$E_{PS} = (3.42 \pm 0.02) \text{ GPa}, \quad (63)$$

$$E_{CF} = (176.66 \pm 2.02) \text{ GPa}. \quad (64)$$

The spring stiffness corresponds to the according longitudinal stiffness:

$$k_s = \frac{E_s A_s}{l_s}, \quad (65)$$

where $A_s$ is the cross sectional area of the rod. However, the stiffness of the sensor used for determination of the spring force has to be taken into account. For the sensor KD24s the spring stiffness is $k_{KD24s} = 2 \cdot 10^7$ N/m. As it is in series with the spring:

$$k_s = \left( \frac{l_s}{E_s A_s} + \frac{1}{k_{KD24s}} \right)^{-1}, \qquad (66)$$

the final stiffness of the springs yield:

$$k_{s-PS} = (2.38 \pm 0.02) \cdot 10^5 \text{ N/m}, \qquad (67)$$

$$k_{s-CF} = (4.34 \pm 0.04) \cdot 10^6 \text{ N/m}. \qquad (68)$$

The stiffness of the sensor TD-112 is not known. As its nominal load is higher than for the KD24s we expect it to be stiffer. Thus, according to (66) its influence is lower. In the end, the crude estimate for the contact stiffness as discussed Appendix A2 leads to the following crude estimates of the parameter $\kappa_1$:

$$\kappa_{1-PS} = \frac{k_{s-PS}}{k_x} \approx 0.1 : \text{used in set 1, 2}, \qquad (69)$$

$$\kappa_{1-SP} = \frac{k_{s-SP}}{k_x} \approx 10^{-4} : \text{used in set 3, 4}, \qquad (70)$$

$$\kappa_{1-CF} = \frac{k_{s-CF}}{k_x} \approx 2 : \text{used in set 5}. \qquad (71)$$

Sets 1-4 are used in the analysis of 4.1 and are tabulated in Table 4. Set 5 is used in the comparison of experiment and simulation in section 4.2 and is described in Table 5.